\documentclass[aps,prl,twocolumn,noshowpacs,superscriptaddress]{revtex4-2}
\usepackage{graphicx,epsfig, subfigure}
\usepackage{amsbsy}
\usepackage{amssymb}
\usepackage{amsmath}
\usepackage{hyperref}
\usepackage[bottom]{footmisc}
\usepackage{xcolor}
\usepackage{color}
\usepackage{times}
\usepackage{textcomp}
\usepackage{verbatim} 

\date{\today}

\begin{document}

\newcommand{\eqnref}[1]{Eq.~\ref{#1}}
\newcommand{\figref}[2][]{Fig.~\ref{#2}#1}
\newcommand{\RN}[1]{%
  \textup{\uppercase\expandafter{\romannumeral#1}}%
}

\title{\Large  Observation of non-contact Casimir friction}

\author{Zhujing Xu}
	\affiliation{Department of Physics and Astronomy, Purdue University, West Lafayette, Indiana 47907, USA}
\author{Peng Ju}
	\affiliation{Department of Physics and Astronomy, Purdue University, West Lafayette, Indiana 47907, USA}
\author{Kunhong Shen}
	\affiliation{Department of Physics and Astronomy, Purdue University, West Lafayette, Indiana 47907, USA}
	\author{Yuanbin Jin}
	\affiliation{Department of Physics and Astronomy, Purdue University, West Lafayette, Indiana 47907, USA}
	\author{Zubin Jacob}
	\affiliation{Elmore Family School of Electrical and Computer Engineering, Purdue University, West Lafayette, Indiana 47907, USA}
	\affiliation{Birck Nanotechnology Center, Purdue University, West Lafayette, Indiana 47907, USA}	
\author{Tongcang Li}
	\email{tcli@purdue.edu}
	\affiliation{Department of Physics and Astronomy, Purdue University, West Lafayette, Indiana 47907, USA}
	\affiliation{Elmore Family School of Electrical and Computer Engineering, Purdue University, West Lafayette, Indiana 47907, USA}
	\affiliation{Birck Nanotechnology Center, Purdue University, West Lafayette, Indiana 47907, USA}	
	\affiliation{Purdue Quantum Science and Engineering Institute, Purdue University, West Lafayette, Indiana 47907, USA}
	\date{\today}

\begin{abstract}
 Quantum mechanics predicts the occurrence of random electromagnetic field fluctuations, or virtual photons, in vacuum. The exchange of virtual photons between two bodies in relative motion could lead to non-contact quantum vacuum friction or Casimir friction. Despite its theoretical significance, the non-contact Casimir frictional force has not been observed and its theoretical predictions have varied widely. In this work, we report the first measurement of the non-contact Casimir frictional force between two moving bodies. By employing two mechanical oscillators with resonant frequencies far lower than those in Lorentz models of electrons in dielectric materials, we have amplified the Casimir frictional force at low relative velocities by several orders of magnitude. We directly measure the non-contact Casimir frictional force between the two oscillators and show its linear dependence on velocity, proving the dissipative nature of Casimir friction. This advancement marks a pivotal contribution to the field of dissipative quantum electrodynamics and enhances our understanding of friction at the nanoscale.
\end{abstract}

\maketitle

The quantum vacuum friction or Casimir friction has been theoretically studied for more than 40 years \cite{reiche2022wading,milton2016reality,10.2307/79496,Schaich_1981,Levitov_1989,Pendry_1997,oue2023noncontact}. After some debates over its existence \cite{Philbin_2009,Pendry_2010,Volokitin_2011}, it is now theoretically believed that two non-contact neutral bodies with a relative motion in vacuum will experience a small dissipative force due to quantum vacuum fluctuations that tend to slow down the relative motion. Although many different schemes have been proposed to detect the non-contact Casimir friction \cite{H_ye_2010,Hoye2011,ge2023negative,fernandez2023spatial,guo2023quantum,PhysRevLett.106.094502,PhysRevLett.109.123604,PhysRevLett.123.120401,Ahn2020,ju2023near,farias2020towards}, 
it has not been observed so far due to experimental challenges \cite{PhysRevLett.83.2402, PhysRevLett.86.2597,Gotsmann2011,reiche2022wading}. Partly due to the lack of experimental results, theoretical predictions of the Casimir friction exhibit considerable variation \cite{milton2016reality,PhysRevA.106.052205}. 
In this work, we report the first measurement of the non-contact Casimir frictional force between two moving objects. Our work is not only important in dissipative quantum electrodynamics, but will also benefit the understanding of friction at the nanoscale \cite{mo2009friction,kavokine2022fluctuation}.

Casimir friction between two surfaces arises due to the interaction between instantaneous charges and electrical dipoles on surfaces induced by quantum fluctuations \cite{Pendry_1997}. 
The dissipation essentially comes from the Doppler shift of the electromagnetic waves reflected by two moving surfaces and the energy eventually dissipates as radiation or heat \cite{Pendry_1997}. 
To detect such small Casimir friction, several studies have proposed to enhance it by resonant photon tunneling \cite{PhysRevLett.91.106101,Guo:14,Guo_2014,XuJacobLi,Volokitin2020,khosravi2024giant}. In this case, the mechanical energy of the moving body can be converted to the electromagnetic energy, leading to an evanescent wave resonance. The resonance requires a critical velocity $V_0$ such that $V_0 = \frac{2\omega_{s}d}{ln|R_p(\omega_{s})|}$ \cite{Guo:14}, where $d$ is the separation between two surfaces, $R_p(\omega) = \frac{\epsilon(\omega)-1}{\epsilon(\omega)+1}$ is the reflection coefficient for p-polarized waves, $\epsilon(\omega)$ is the dielectric function of the material, and $\omega_{s}$ is the surface wave resonance frequency such that $Re(\epsilon(\omega_s)) = -1$. 
However, this critical velocity is extremely high for conventional dielectric materials and  makes the measurement extremely difficult.  For a typical dielectric material such as silicon carbide (SiC), the surface wave resonance frequency is $\omega_s = 1.76\times 10^{14}$~rad/s \cite{PhysRevLett.118.133605}.  At a separation of 100~nm, the resonant 
velocity is $1.31\times 10^{7}$~m/s. Such an extremely high required velocity suppresses the possibility of measuring the Casimir friction with a conventional material. 

\begin{figure*}
	\centerline{\includegraphics[width=1.0\linewidth]{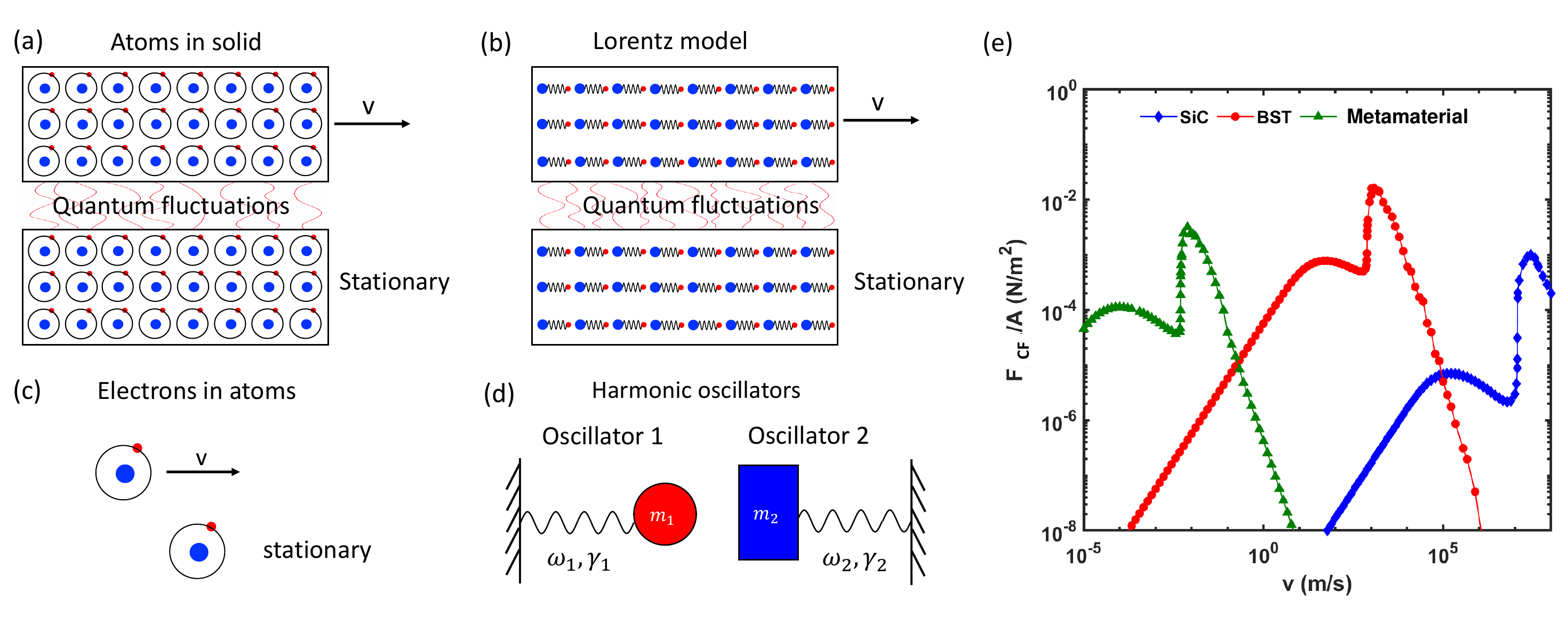}}
	\caption{\textbf{Casimir friction described by different models.} (a). When two closely spaced solid plates have a relative motion, they will experience the Casimir frictional force against the moving direction. The Casimir friction force essentially comes from the interaction between the incident photons and the electrons in solids. (b). The optical properties of solids can be  modeled as Lorentz oscillators, which treat each	discrete vibrational mode as a classical damped harmonic oscillator. The Casimir friction force here comes form the interaction between the incident photons and the phonons in solids. (c). In a microscopic point of view, the friction force exists between two moving atoms similar to the mechanism in Fig.(a). (d). In a microscopic point of view, the friction force exists between two classical damped oscillators with a relative motion, similar to the mechanism in Fig.(b). (e). Calculated Casimir frictional force between two moving dielectric materials at a separation of 100 nm. We notice that the resonant velocity are around $1\times 10^7$ m/s, $1\times 10^3$ m/s and $1\times 10^{-2}$ m/s for SiC, BST and a hypothetical metamaterial.}
	\label{Schematic}
\end{figure*}

If we can decrease the surface wave resonance frequency, we can greatly reduce the challenge of detecting the non-contact Casimir frictional force.  
To demonstrate this, we show the calculation of Casimir friction ($F_{CF}$) due to quantum vacuum fluctuations  for different dielectric materials in Fig.\ref{Schematic}(a), where one solid plate is moving with a velocity $v$ while the other solid plate is stationary\cite{Volokitin_1999}. The friction force origins from the interaction between the incident virtual photons (quantum vacuum fluctuations) and the electrons in solids (Fig.\ref{Schematic}(a)).
For simplicity, here we assume two separated plates in relative parallel motion are made of identical materials \cite{PhysRevLett.91.106101}. 
The Casimir friction depends on the material properties, the separation between two plates and the relative velocity. The material properties can be described by the dielectric functions and the dielectric functions of plates are modeled by Lorentz models, which treat each discrete vibrational mode as a classical damped harmonic oscillator as shown in Fig.\ref{Schematic}(b). More details are included in the Supplementary Material Section \textrm{II}. In this case, the energy dissipates through the phonon loss of the vibrational modes in solids.

Figure \ref{Schematic}(e) shows the calculation results for SiC \cite{PhysRevLett.118.133605}, barium strontium titanate (BST) \cite{XuJacobLi} and a hypothetical metamaterial. The metamaterial is assumed to have a Lorentz oscillator resonant frequency of $2\pi\times 5$ kHz and a damping rate of $2\pi\times 100$ Hz. 
The surface wave resonance frequencies are $2\pi\times 2.80\times 10^{13}$ Hz, $2\pi\times1.84\times 10^9$ Hz, $2\pi\times 1.18\times 10^4$ Hz for SiC, BST and the metamaterial, respectively. 
Therefore, the critical velocities for resonant photon tunneling at a separation of 100 nm are around $1.3\times 10^7 $ m/s, $840$ m/s and $5.2\times 10^{-3}$ m/s, respectively. 
In Fig.\ref{Schematic}(e), we show that the resonant velocity for the metamaterial is more than nine orders lower than the resonant velocity of SiC. This leads to an enhancement of Casimir friction coefficient $F_{CF}/v$ by more than nine orders. 
Therefore, a metamaterial with a resonant frequency at kHz can significantly enhance the Casimir friction coefficient and makes the detection feasible. A normal relative motion also gives an enhancement of Casimir friction force compared to the parallel motion case \cite{PhysRevLett.91.106101}.

Inspired by results in Fig.\ref{Schematic}(e), we develop a novel way to enhance the Casimir friction force at low relative velocities.  
When two solids have a relative motion, the quantum vacuum  fluctuations between them induce the Casimir friction force (Fig.\ref{Schematic}(a)).
This Casimir friction force essentially comes from the interaction between virtual photons, electrons and nuclei in the solids.   The optical properties of macroscopic dielectric materials can be well modeled by Lorentz oscillators (Fig.\ref{Schematic}.(b)).
Phonons (nuclear vibrations) play a critical role in the Casimir friction force at low velocities (Fig.\ref{Schematic}(e)). 
Casimir friction can also exist at a microscopic scale. Two moving atoms (Fig.\ref{Schematic}.(c))) will also experience Casimir friction due to quantum vacuum fluctuations, which has been calculated by modeling the two atoms as moving harmonic oscillators \cite{H_ye_2010,Hoye2011,HOYE2012}. 
Similar to the two-atom system, a system consists of two moving mechanical oscillators will also experience the Casimir friction (Fig.\ref{Schematic}.(d)).

In this work, we measure the non-contact Casimir friction between two mechanical harmonic oscillators with relative motion as shown in Fig.\ref{Schematic}(d). 
The resonant frequencies of the mechanical resonators are about 5 kHz, similar to the metamaterial discussed above.  
This experimental scheme significantly reduces the requirement of the relative moving velocity. 
For solids, the system energy is eventually dissipated by the electrical resistance and phonon damping in the solids. For mechanical harmonic oscillators, the system energy is dissipated by the mechanical loss of the resonators. 
We exploit a unique approach to minimize the non-dissipative Casimir force such that the frictional component can be isolated through a direct measurement. 
Besides, we can engineer the system loss to maximize the Casimir frictional force. This provides flexibility of tuning the Casimir frictional force. 
In this manuscript, we will show that we are able to increase a Casimir frictional force up to more than $3$ pN at $0.38$ mm$/$s.  This is the first observation of Casimir frictional force. We show the linear velocity dependence of this dissipative Casimir friction force, revealing the dissipative nature of the frictional force.
We also experimentally observed the quantum vacuum induced damping coefficient and isolated the instantaneous friction force.

\begin{figure}
	\centerline{\includegraphics[width=1\linewidth]{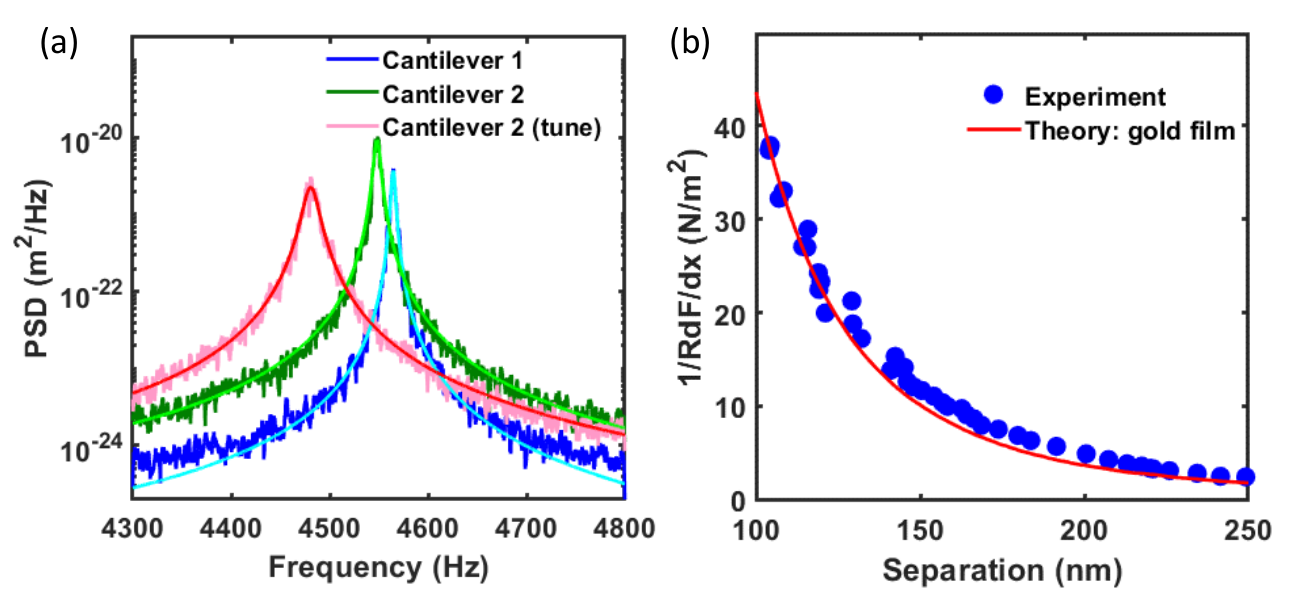}}
	\caption{\textbf{Measurement of Casimir interaction between two cantilevers.} (a). Power spectrum density (PSD) of cantilever 1 and cantilever 2. The blue one shows the PSD of cantilever 1 with a natural frequency of 4564.8 Hz. The green one shows the PSD of cantilever 2 with a natural frequency of 4548.9 Hz. In the Casimir force measurement, the frequency of cantilever 2 is shifted to 4481.5 Hz by PID control, in order to prevent resonant coupling, as shown in the red curve. (b). Measured Casimir force gradient is shown as a function of separation. } 
	\label{Casimirforce}
\end{figure}

\begin{figure*}
	\centerline{\includegraphics[width=1\linewidth]{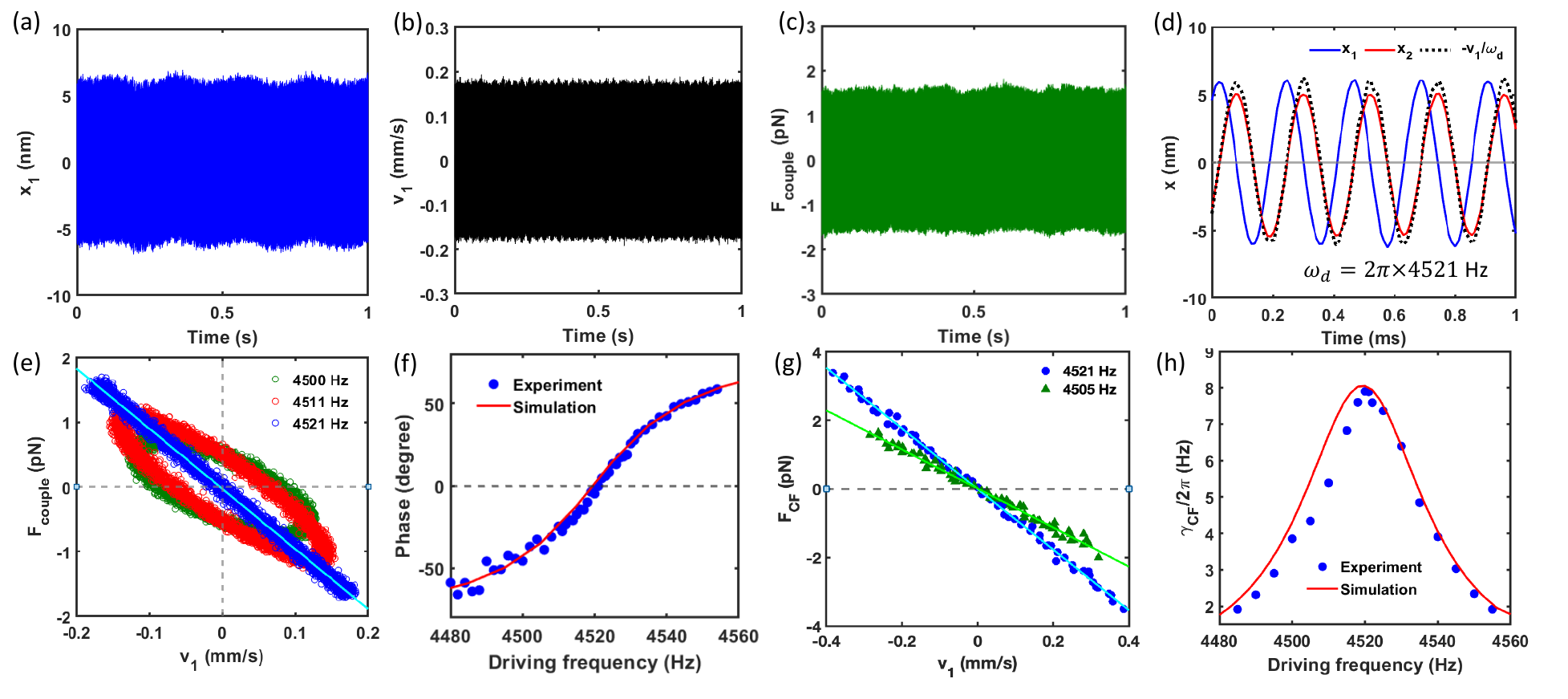}}
	\caption{\textbf{Measurement of Casimir friction force.} (a)-(c). Recorded $x_1$, $v_1$ and $F_{couple} =  |\frac{dF_{Casimir}}{dx}|x_2$ under the driving of 4521 Hz on cantilever 1 when the separation is 154 nm. Here the driving frequency matches the resonant frequency of cantilever 2. (d).  $v_1$ and $x_2$ are in phase, revealing that the coupling force is fully disspative. (e). The measured coupling force is shown for different driving frequencies and different velocities. When the driving frequency matches the resonant frequency of cantilever 2 such that $\omega_d = \omega_2'$, the coupling force $F_{couple}$  is linear with the velocity $v_1$ and hence is a fully dissipative force. When the driving frequency has a detuning from resonance, the coupling force has a non-zero conservative component.  (f). The relative phase between $-v_1$ and $F_{couple}$. It is noticed that for a resonant case, the phase becomes zero. (g). The Casimir friction force $F_{CF}$ at each velocity is extracted from $F_{couple}$ when $x_1$ equals to 0.  (h). The Casimir friction damping rate $\gamma_{CF} = F_{CF}/m_1v_1$ is shown for different driving frequency $\omega_d$.} 
	\label{Forcemeasure}
\end{figure*}

Experimentally, our mechanical harmonic oscillators are two modified atomic force microscope (AFM) cantilevers which have nearly the same resonant frequencies to maximize the Casimir friction. A 70-$\mu$m-diameter microsphere is attached to the free end of one cantilever \cite{xu2022non}. Both the microsphere and cantilever surfaces are coated with  100-nm-thick gold films. The gap between the microsphere and the flat surface of the other cantilever is on the order of 100 nm. They are coupled by the Casimir force \cite{PhysRev.73.360,lamoreaux1997demonstration,xu2022non,xu2022observation,zhao2019stable,munkhbat2021tunable,pate2020casimir} due to quantum vacuum fluctuations. More details of the dual-cantilever system can be found in Supplementary Material Section  \textrm{I}.
The power spectrum density (PSD) of two cantilevers are shown in Fig.\ref{Casimirforce}.(a). The natural frequency of two cantilevers are $\omega_1 = 2\pi\times 4564.7$ Hz and $\omega_2 = 2\pi\times 4548.9$ Hz, respectively. The resonant frequency can be tuned by an external PID feedback. In order to measure the Casimir force without resonant coupling, the frequency of cantilever  2 is shifted to $\omega_2' = 2\pi\times 4481.5$ Hz as shown in the red curve in Fig.\ref{Casimirforce}(a).   We measure the Casimir force gradient $\frac{1}{R}\frac{dF}{dx}$ from the motion of cantilever 2 with the off-resonant frequency. The experimental results are shown in Fig.\ref{Casimirforce}.(b), which agree well with theoretical calculations (see Supplementary Material Section \textrm{III}) \cite{Lifshitz:1956,BLOCKI1977427}. Notice that the electrostatic force between two cantilevers due to surface patch potentials is cancelled out by applying additional voltages, similar to the method in \cite{xu2022non}. Therefore, the coupling force between two cantilever is purely from Casimir interactions. 
Besides, the thermal contribution of the Casimir force is about $2\%$ when the separation is less than 200 nm and when the system is at room temperature\cite{xu2022non}. 
This unique dual-cantilever system has been used previously to demonstrate the Casimir effect as well as a Casimir diode\cite{xu2022non}. However, the dissipative Casimir frictional force has not been measured before. 

The measurement of the Casimir friction force is demonstrated as follows. 
The Casimir force between the two cantilevers is separation-dependent and sensitive to the motion of two cantilevers. Under the small vibration amplitube approximation, the Casimir force on cantilever 1 is written as $F_{Casimir} = 	F_{Casimir}(d)-\frac{dF_{Casimir}}{dx}(d)(x_1+x_2)$. Here we focus on the coupling force on cantilever 1 which is $F_{couple}= -\frac{dF_{Casimir}}{dx}(d)x_2$.
The Casimir coupling force between two cantilevers can be separated into the conservative term and the dissipative term as follows, 
\begin{equation}
	F_{couple} = F_{conservative}+F_{CF},
\end{equation}
where the dissipative force corresponds to the Casimir friction force $F_{CF}$. Under a sinusoidal driving force on cantilever 1 with driving frequency $\omega_d$, the two parts of the coupling force at the steady state are
\begin{eqnarray}
	F_{conservative} = \frac{(dF_{Casimir}/dx)^2}{m_2}\frac{\omega_2'^2-\omega_{d}^2}{(\omega_2'^2-\omega_{d}^2)^2+\gamma_2^2\omega_{d}^2}x_1,\nonumber\\
	F_{CF} = -\frac{(dF_{Casimir}/dx)^2}{m_2}\frac{\gamma_2}{(\omega_2'^2-\omega_{d}^2)^2+\gamma_2^2\omega_{d}^2}\dot{x}_1,
	\label{QVF}
\end{eqnarray}
where $\omega_1' = \omega_1\sqrt{1+\frac{1}{m_1 \omega_1^2}\frac{dF_{Casimir}}{dx}}$, $\omega_2' = \omega_2\sqrt{1+\frac{1}{m_2 \omega_2^2}\frac{dF_{Casimir}}{dx}}$ and they are the resonant frequency of two cantilevers in the presence of the Casimir force $F_{Casimir}$.
$F_{conservative}$ is the conservative term which only depends on $x_1$ and $F_{CF}$ is the dissipative term which only depends on $\dot{x}_1$. 
More details about the derivation can be found in the Supplementary Material Section \textrm{III}.
Here we consider a special case that when the driving frequency equals to the resonant frequency of cantilever 2 such that $\omega_{d} = \omega_{2}'$, the conservative force $F_{conservative}$  will be zero and the coupling force $F_{couple}$ equals to the Casimir friction force $F_{CF}$. In this way, we can minimize the non-dissipative component of the coupling force and directly measure the Casimir friction. 

In this experiment, we measure the instantaneous Casimir friction force and the results are shown in Fig.\ref{Forcemeasure}. Here the damping rate of two cantilevers are tuned by the PID feedback loop to be $\gamma_1 = 2\pi\times 3.7$ Hz and $\gamma_2 = 2\pi\times 41.6$  Hz. The tuning details can be found in the Supplementary Material Section \textrm{IV}.
 Two cantilevers are closely spaced by a separation of 154 nm. Under a driving of 4521 Hz on cantilever 1, we record the instantaneous displacements and velocities of two cantilevers. Here the resonant frequency $\omega_2'$ of cantilever 2 at the presence of  Casimir force is 4521 Hz.
 The time-dependent displacement $x_1$ and velocity $v_1$ of cantilever 1  are measured by the fiber interferometer and are shown in Fig.\ref{Forcemeasure}.(a) and (b).  Since the coupling force on cantilever 1 depends on the displacements of cantilever 2.
 We could measure $x_2$ and the force gradient $\frac{dF_{Casimir}}{dx}$ and get the instantaneous Casimir coupling force $F_{couple}$  as shown in Fig.\ref{Forcemeasure}.(c).
The separation here is 154 nm. 
We notice that this Casimir coupling force is time varying, which eventually dissipates the energy by the mechanical loss of the cantilevers.

To further understand the distribution of conservative and dissipative terms in the coupling force, we show the recorded $x_1$, $x_2$ and $-v_1/\omega_d$ in a short period of time in  Fig.\ref{Forcemeasure}.(d). When the driving frequency matches the resonant frequency of cantilever 2 such that $\omega_d = \omega_2'$, $x_2$ and $-v_1$ are in phase and the coupling force is linear with $v_1$,  revealing that it is a fully dissipative force and this is the Casimir friction force $F_{CF}$. This agrees with the conclusion in Eq.\ref{QVF}. We can easily plot the relation between $F_{couple}$ and $v_1$ in Fig.\ref{Forcemeasure}.(e) for different velocities. In addition to the resonant driving of 4521 Hz (the driving frequency matches the resonant frequency of cantilever 2), we also present two off-resonant cases with a detuning of -10 Hz and -21 Hz. For the resonant case, $F_{couple}$ and $v_1$ shows a linear relation because $F_{couple}$ only comes from the dissipative force $F_{CF}$. For the off-resonant case, it shows an elliptical relation, depending on the detuning of the driving frequency. It indicates that the coupling force has a non-zero conservative component. The phase between $-v_1$ and $F_{couple}$ is shown in Fig.\ref{Forcemeasure}.(f). This is the first observation of Casimir friction and the heat dissipates as the mechanical loss of the resonators. 
Here two mechanical resonators have unique relative motion around their center-of-mass, which is different from the original vacuum friction problem with a constant relative velocity but has important similarities \cite{10.2307/79496,Schaich_1981,Levitov_1989,Pendry_1997}.

The key signature of the measured effect is that the Casimir friction force depends on the relative velocity. To understand the relation between the Casimir friction force and the velocity, we measure the friction force $F_{CF}$ at different velocities by extracting the time-dependent $v_1$ and $x_2$ at a group of specific time when  $x_1 = 0$. 
Under such condition, the conservative component of the coupling force in Eq.\ref{QVF} becomes zero so we can have $F_{couple} = F_{CF}$.  The recorded $F_{CF}$ is shown in Fig.\ref{Forcemeasure}.(g). 
The measured Casimir friction force can reach up to more than $3$ pN at a velocity around $4\times 10^{-4}$ m/s for a resonant case. 
A off-resonant driving will give a smaller friction force $F_{CF}$ for the same velocity as shown in Fig.\ref{Forcemeasure}.(g). 
A linear fitting gives the Casimir friction damping rate $\gamma_{CF} = F_{CF}/m_1v_1$ for each driving frequency.  
We notice that, the measured Casimir friction damping rate can reach up to $2\pi\times 7.9$ Hz at a separation of 154 nm. 
Considering an effective interaction area of $A = \frac{2}{3}\pi Rd = 1.13\times 10^{-11}$ m$^2$, we are able to detect a Casimir friction stress $\sigma_{CF} = F_{CF}/A = 0.3$ N$/$m$^2$ and a corresponding Casimir friction coefficient of $\Gamma_{CF} = F_{CF}/v_1A = 783$ kgs$^{-1} m^{-2}$.
A larger Casimir friction coefficient is expected to be realized at a smaller separation.  This method significantly enhances the friction force and lowers the required velocity. As a comparison, the Casimir friction force coefficient between two moving SiC plates with a separation of 1 nm is calculated to be about $0.014$ kgs$^{-1} m^{-2}$\cite{PhysRevLett.91.106101}. Our experiment shows an enhancement of Casimir friction coefficient by several orders of magnitudes compared to the conventional scheme.

\begin{figure}	\centerline{\includegraphics[width=1\linewidth]{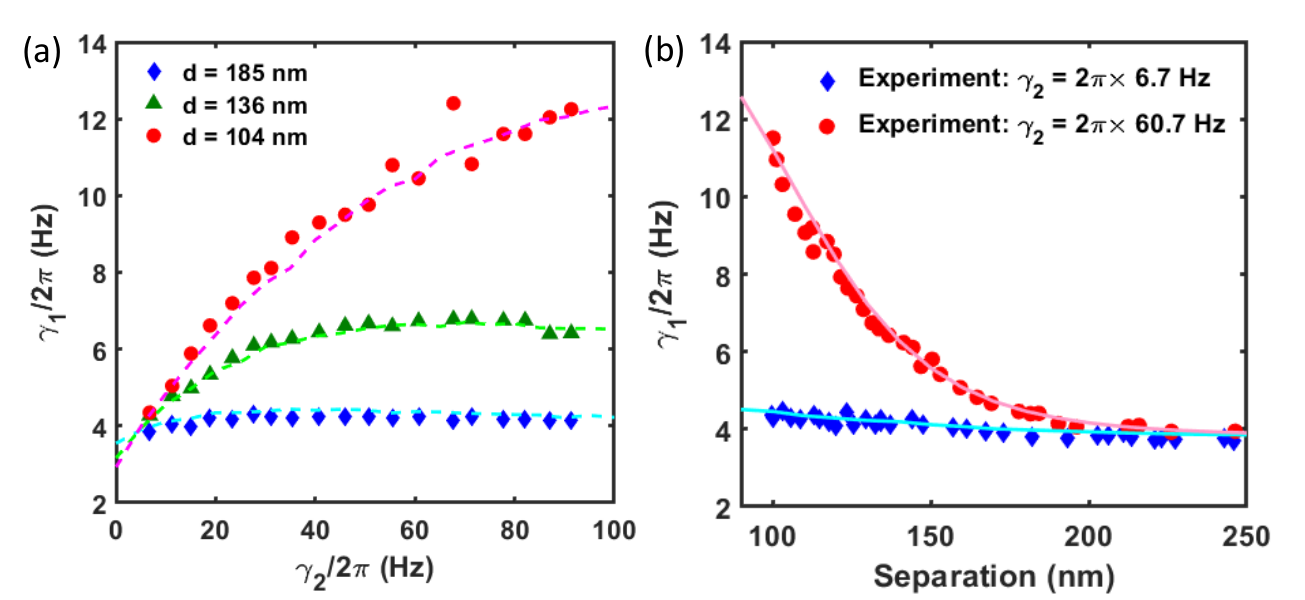}}
	\caption{\textbf{Measured mechanical loss through quantum vacuum fluctuations.}  (a). The measured damping rate of cantilever 1 $\gamma_1$ is shown as a function of a tunable $\gamma_2$. A higher mechanical loss $\gamma_2$ gives a larger $\gamma_1$ because of a stronger Casimir friction. The dashed curves correspond to the simulations. (b). The measured $\gamma_1$ is shown as a function of the separation. }
	\label{Dampmeasure}
\end{figure}

After measuring the Casimir friction force between two cantilevers, we measured the damping rate of cantilever 1 when two cantilevers are closely spaced.  When cantilever 2 has a high mechanical loss $\gamma_2$, cantilever 1 is expected to experience a large Casimir friction force and hence the damping rate $\gamma_1$ should increase. 
The measured $\gamma_1$ is shown in Fig.\ref{Dampmeasure}.(a) and (b) for different $\gamma_2$ and separations. The damping rate is extracted by fitting the frequency response of cantilever 1 with a Lorentzian function.
 We notice that $\gamma_1$ increases as $\gamma_2$ increases and then saturated, eventually limited by the coupling strength between two cantilevers. The damping rate $\gamma_1$ can reach up to about $2\pi\times 12$ Hz at a separation of 99 nm. The separation dependence of $\gamma_1$ is shown in Fig.\ref{Dampmeasure}.(b). In this case, the damping rate $\gamma_1$ of cantilever 1  reflects the energy loss by cantilever 1 mediated by quantum vacuum fluctuations.

In conclusion, we have experimentally observed the Casimir friction force in a mechanical system. The  mechanism of friction force essentially comes from an  exchange of particles \cite{Pendry_1997}. In our case, the Casimir friction comes from the exchange of virtual photons after interacting with the phonons in the mechanical system. 
The friction dissipates energy through the mechanical loss mediated by the Casimir force. 
Experimentally, we measure the instantaneous Casimir friction force up to more than $3$ pN and demonstrate its linear dependence on velocity. To our best knowledge, this is the first measurement of Casimir friction force. Our detection of Casimir friction force will provide a better understanding of non-contact friction at the nanoscale \cite{mo2009friction,kavokine2022fluctuation}, and will be the precursor to  observe the quantum vacuum friction under other conditions \cite{H_ye_2010,Hoye2011,ge2023negative,fernandez2023spatial,guo2023quantum,PhysRevLett.106.094502,PhysRevLett.109.123604,PhysRevLett.123.120401,Ahn2020,ju2023near,farias2020towards}. 
Since the non-contact friction applies to a lot of scenarios, our work will benefit the micro and nanomechanical technology as well as mechanical detection and sensors \cite{Gotsmann2011}.









%


\newpage
\onecolumngrid
\appendix

\section*{Supplementary Material}

\section{Experimental setup and force measurement}

\begin{figure}[h]
	\centerline{\includegraphics[width=0.6\linewidth]{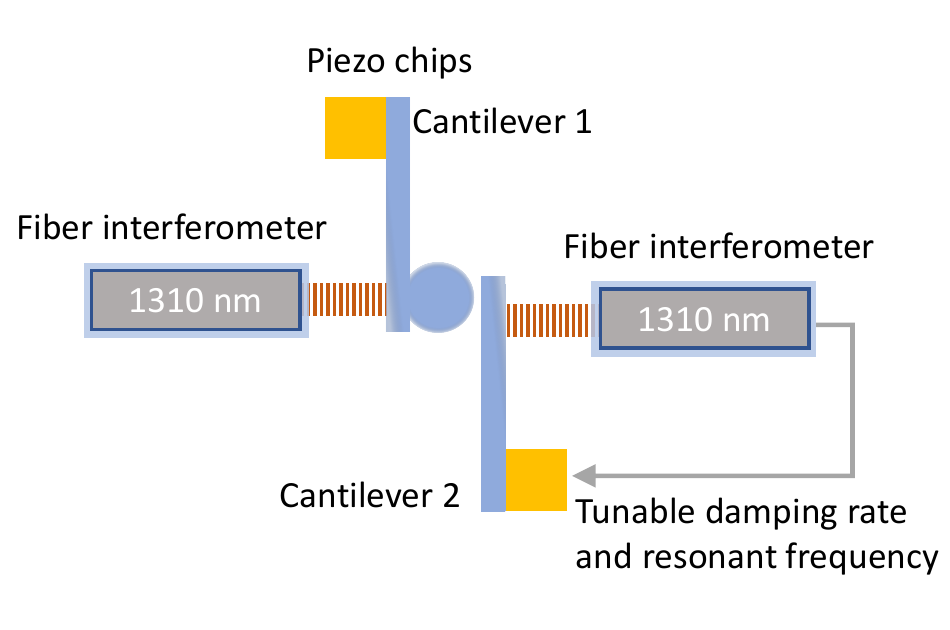}}
	\caption{ The schematic of the Casimir friction measurement setup. Two modified AFM cantilevers are used to detect the dissipative Casimir friction force. Two fiber interferometers are implemented to measure the motions of the two cantilevers. The piezo chips at the end of the cantilevers are used to control the motion of the cantilevers. A PID feedback control loop is applied to cantilever 2 through the piezo chips to tune its damping rate and resonant frequency. }
	\label{Scheme_setup}
\end{figure}

The Casimir friction measurement system consists of two modified atomic force microscope (AFM) cantilevers as shown in Fig.\ref{Scheme_setup}, similar to the setup in \cite{xu2022non}. The left cantilever has a dimension of 450 $\mu$m $\times$ 50 $\mu$m $\times$ 2 $\mu$m. A $70-$$\mu$m-diameter
polystyrene sphere is attached to the free end of the left cantilever. 
The right cantilever has a dimension of 500 $\mu$m $\times$ 100 $\mu$m $\times$ 1 $\mu$m. Additional 100-nm-thick gold layers are coated on both the sphere and cantilever surfaces to ensure good conductivity.  Two piezo chips are mounted at the end of the cantilevers to control the motion. 
In the Casimir friction measurement, two fiber interferometers with a 1310-nm-wavelength laser are implemented to measure the instantaneous motion and velocity of two cantilevers. Besides, a PID feedback control loop is applied to cantilever 2 through the piezo chips to tune its damping rate and resonant frequency. More details of the tunable damping rate can be found in section \ref{section_damp}.

\section{Calculation of Casimir friction between two sliding plates}

\begin{figure}	\centerline{\includegraphics[width=0.45\linewidth]{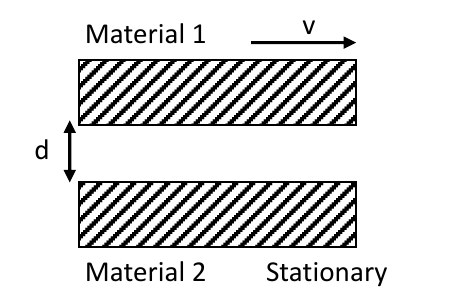}}
	\caption{ Two perfectly smooth and infinitely thick plates are separated by a distance $d$ and they are having a relative motion with a velocity $v$. They experience a friction force against the moving direction because of the quantum vacuum fluctuations. }
	\label{Scheme_QVF}
\end{figure}

We consider two perfectly smooth plates moving with a relative parallel motion. The schematic is shown in Fig.\ref{Scheme_QVF}. In the nonrelativistic limit ($v\ll c$) and
near-field limit ($d\ll c\hbar/k_BT$), the Casimir friction force per unit area experienced by the plates is \cite{Volokitin_1999} 
\begin{eqnarray}
	F_{CF} = \frac{\hbar}{4\pi^3}\int_{-\infty}^{\infty}dq_x\int_{-\infty}^{\infty}dq_y\int_{0}^{\infty}d\omega q_x 
	\{\{\frac{Im(R_{1p}(\omega))Im(R_{2p}(\omega-q_xv))e^{-2qd}}{|1-e^{-2qd}R_{1p}(\omega)R_{2p}(\omega-q_xv)|^2}+(p\leftrightarrow s)\}\nonumber\\
	\times[n_2(\omega-q_xv)-n_1(\omega)]+(1\leftrightarrow 2)\},\hspace{0.5cm}
\end{eqnarray}
where $v$ is the relative velocity between two plates, $n_{1,2}(\omega) = \frac{1}{exp(\hbar\omega/k_BT_j)-1}$ is the Bose–Einstein
distribution function at temperature $T_j$ and $j = 1,2$ are for two different plates. $R_{jp(s)}$ is the reflection amplitude of plate $j$ for p and s polarized electromagnetic waves and it is written as 
\begin{eqnarray}
	R_{jp} = \frac{\epsilon_j p-s_j}{\epsilon_j p+s_j}, \hspace{0.5cm} 
	R_{js} =  \frac{\epsilon_j -s_j}{\epsilon_j +s_j}.\hspace{3.8cm} \nonumber\\
	q = \sqrt{q_x^2+q_y^2}, \hspace{0.5cm} p = \sqrt{\frac{\omega^2}{c^2}-q^2}, \hspace{0.5cm} s_j = \sqrt{\frac{\omega^2}{c^2}\epsilon_j-q^2}.
\end{eqnarray}

\begin{figure}	\centerline{\includegraphics[width=1\linewidth]{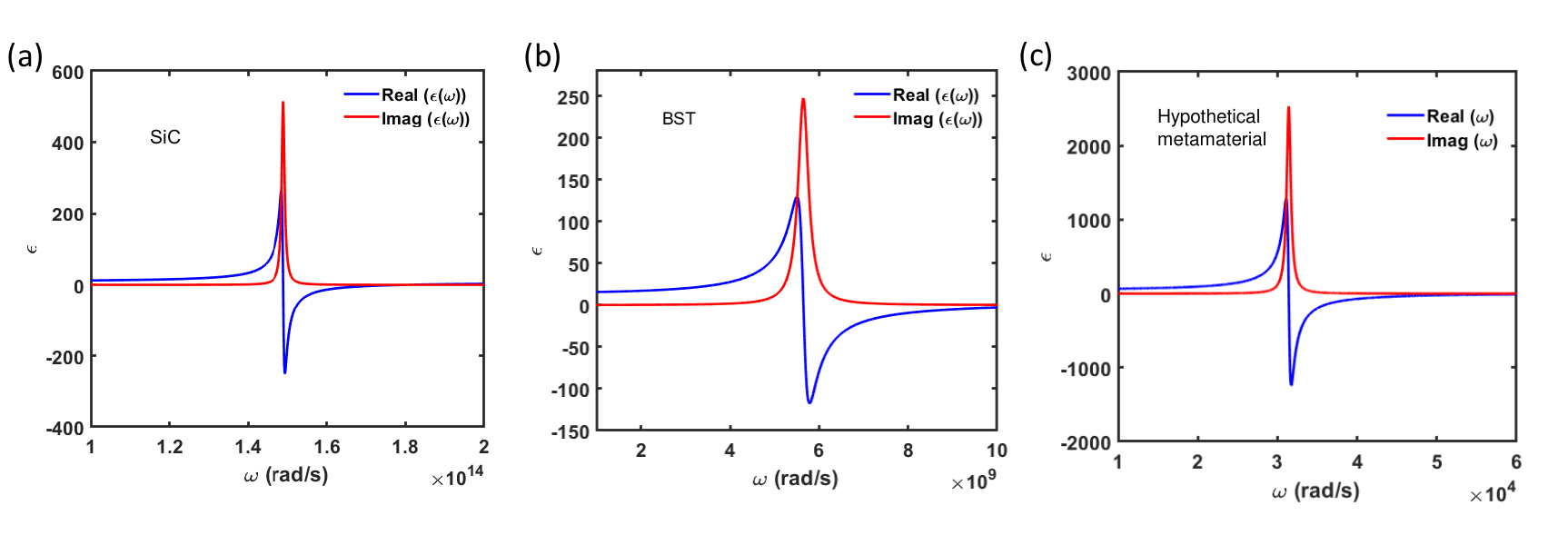}}
	\caption{ The dielectric function for three different materials. }
	\label{Dielectric}
\end{figure}

Here we calculate the Casimir friction force between two sliding plates at a separation of $100$~nm. The temperature of both plates are at $300$~K. The calculated friction force is shown in Fig.1.(e) in the main text.  The dielectric functions of different dielectric materials can be modeled as Lorentz oscillators, which treat each discrete vibrational mode as a classical damped harmonic oscillator. The dielectric functions  can be written as 
\begin{equation}
	\epsilon(\omega) = \epsilon_{\infty} (1+\frac{\omega_L^2-\omega_T^2}{\omega_T^2-\omega^2-i\gamma\omega}).
\end{equation}
The parameters for SiC are $\epsilon = 6.7$, $\omega_L = 1.8\times  10^{14}$ s$^{-1}$,   $\omega_T  = 1.5\times 10^{14}$ s$^{-1}$ and $\gamma = 8.9\times 10^{11}$ s$^{-1}$\cite{PhysRevLett.118.133605}. The parameters for barium barium strontium titanate (BST) are $\epsilon_{\infty} = 2.9$, $\omega_L = 1.3\times  10^{10}$ s$^{-1}$,   $\omega_T  = 5.7\times 10^{9}$ s$^{-1}$ and $\gamma = 2.8\times 10^{8}$ s$^{-1}$\cite{XuJacobLi}. The real and imaginary part of the dielectric function for these two materials are shown in Fig.\ref{Dielectric}.(a) and (b). In addition, we investigate a hypothetical metamaterial that has a resonant frequency at 5 kHz and a dielectric function that can be described by the Lorentz model: 
\begin{equation}
	\epsilon(\omega) = \epsilon_{\infty}(1+\frac{B}{\omega_T^2-\omega^2-i\gamma\omega}),
\end{equation} 
where the parameters are engineered to be $\omega_T  = 2\pi \times 5000$ s$^{-1}$, $\gamma = 2\pi\times 100$ s$^{-1}$, $B = 5\times 10^9$ s$^{-2}$.

\section{Casimir friction force between two harmonic oscillators}

In this part, we will show the equations of motion in details and derive the dissipative Casimir friction force. The equation of motion of the two-cantilever system can be written as 
\begin{eqnarray}
	m_1\ddot{x_1}+m_1\gamma_1\ddot{x_1}+m_1\omega_1^2 x_1 = F_{Casimir}(t)+ F_{d}(t),\nonumber\\
	m_2\ddot{x_2}+m_2\gamma_2\ddot{x_2}+m_2\omega_2^2 x_2 = F_{Casimir}(t),\hspace{1.3cm}
\end{eqnarray}
where the driving force is $F_{d} = F_0\cos(\omega_{d}t)$ with driving amplitude $F_0$ and driving frequency $\omega_{d}$. The Casimir force here under the proximity-force approximation (PFA)\cite{BLOCKI1977427} is given by 
\begin{equation}
	F_{Casimir}(x,T) = -2\pi RE(x,T),
\end{equation}  
where $E(x,T)$ is the Casimir pressure between two infinitely thick plates. At a finite temperature and at a separation $x$, the Casimir pressure $E(x,T)$ is\cite{Lifshitz:1956} 
\begin{equation}
	E(x,T)	= \frac{k_BT}{2\pi}\sum_{l=0}^{\infty} \prime\int_0^{\infty}k_{\perp}dk_{\perp}\{\ln[1-r_{TM}^{2}(i\xi_l,k_{\perp})e^{-2xq}]+\ln[1-r_{TE}^{2}(i\xi_l,k_{\perp})e^{-2xq}]\}
\end{equation}
where $\xi_l = \frac{2\pi k_BTl}{\hbar}$ is the Matsubara frequency and the prime on the summation indicates that the $l = 0$ term will be multiplied by $1/2$. We notice that the Casimir  force is separation-dependent and very sensitive to the oscillation of two cantilevers so the time-dependent Casimir force is written as $F_{Casimir}(t) = F_{Casimir}(d-x_1(t)-x_2(t))$.
When the oscillation of two cantilevers near the equilibrium are far smaller than the separation, the Casimir force can be expanded to the first order and be written as 
\begin{eqnarray}
	F_{Casimir} (d-x_1-x_2) = F_{Casimir}(d)-\frac{dF_{Casimir}}{dx}(d)(x_1+x_2).
\end{eqnarray}
Therefore, the equations of motion will become
\begin{eqnarray}
	m_1\ddot{x_1}+m_1\gamma_1\ddot{x_1}+m_1\omega_1'^2 x_1 = Jx_2+F_0\cos(\omega_{d}t),\nonumber\\
	m_2\ddot{x_2}+m_2\gamma_2\ddot{x_2}+m_2\omega_2'^2 x_2 = Jx_1.\hspace{2.2cm}
\end{eqnarray}
where $\omega_1' = \omega_1\sqrt{1+\frac{1}{k_1}\frac{dF_{Casimir}}{dx}}$, $\omega_2' = \omega_2\sqrt{1+\frac{1}{k_2}\frac{dF_{Casimir}}{dx}}$ and $J = -\frac{dF_{Casimir}}{dx}(d)$. $J$ is a positive value since the Casimir force in our vacuum system is an attractive interaction.  
Here we focus on the coupling force on cantilever 1 which is $F_{couple} = Jx_2$. The general solutions for the above equations are
\begin{eqnarray}
	x_1 = \frac{(\omega_2'^2-\omega_{d}^2+i\gamma_2\omega_{d})f}{(\omega_1'^2-\omega_{d}^2+i\gamma_1\omega_{d})(\omega_2'^2-\omega_{d}^2+i\gamma_2\omega_{d})-j^2}\times exp(i\omega_{d}t),\nonumber\\
	x_2 = \sqrt{\frac{m_1}{m_2}}\frac{jf}{(\omega_1'^2-\omega_{d}^2+i\gamma_1\omega_{d})(\omega_2'^2-\omega_{d}^2+i\gamma_2\Omega)-j^2}\times exp(i\omega_{d}t),
\end{eqnarray}
where we have $j = J/\sqrt{m_1m_2}$ and $f = F/m_1$. When the driving frequency matches the resonant frequency of cantilever 2 such that $\omega_{d} = \omega_2'$, the solutions will become
\begin{eqnarray}
	x_1 = \frac{i\gamma_2\omega_2'f}{(\omega_1'^2-\omega_2'^2+i\gamma_1\omega_2')i\gamma_2\omega_2'-j^2} exp(i\omega_2't),\hspace{1cm}\nonumber\\
	x_2 = \sqrt{\frac{m_1}{m_2}}\frac{jf}{(\omega_1'^2-\omega_2'^2+i\gamma_1\omega_2')i\gamma_2\omega_2'-j^2}exp(i\omega_2't).\hspace{0.6cm}
\end{eqnarray}
Under such case, the relation between $x_1$ and $x_2$ will be 
\begin{equation}
	x_2 = -i\sqrt{\frac{m_1}{m_2}}\frac{j}{\gamma_2\omega_2'}x_1 = -\frac{J}{m_2\gamma_2\omega_2'^2}\dot{x_1},
\end{equation}
which means that we can rewrite the coupling term on cantilever 1 such that 
\begin{equation}
	F_{couple} = Jx_2 = -\frac{J^2}{m_2\gamma_2\omega_{d}^2}\dot{x_1}.
\end{equation}
Therefore, for a special driving that $\omega_{d} = \omega_2'$, the coupling force all comes from the dissipative Casimir friction force. There is no conservative part under such condition. 
A more general case of the coupling force, the conservative force and the dissipative friction force will be 
\begin{eqnarray}
	F_{couple} = \frac{J^2}{m_2}(\frac{\omega_2'^2-\omega_{d}^2}{(\omega_2'^2-\omega_{d}^2)^2+\gamma_2^2\omega_{d}^2}x_1
	-\frac{\gamma_2}{(\omega_2'^2-\omega_{d}^2)^2+\gamma_2^2\omega_{d}^2}\dot{x_1}),\nonumber\\
	F_{conservative} = \frac{J^2}{m_2}\frac{\omega_2'^2-\omega_{d}^2}{(\omega_2'^2-\omega_{d}^2)^2+\gamma_2^2\omega_{d}^2}x_1,\hspace{3.7cm}\nonumber\\
	F_{CF} = -\frac{J^2}{m_2}\frac{\gamma_2}{(\omega_2'^2-\omega_{d}^2)^2+\gamma_2^2\omega_{d}^2}\dot{x_1},\hspace{4.5cm}
\end{eqnarray}
where the conservative part $F_{conservative}$ only depends on $x_1$ and the dissipative part $F_{CF}$ only depends on $\dot{x_1}$.
The main idea of this paper is to measure the dissipative Casimir friction force $F_{CF}$ mediated by the quantum vacuum fluctuations and  its dependence of the mechanical intrinsic loss of the system. 

\section{Calibration of the tunable damping rate of the harmonic resonators} \label{section_damp}

\begin{figure}[tb]	
	\centerline{\includegraphics[width=1.0\linewidth]{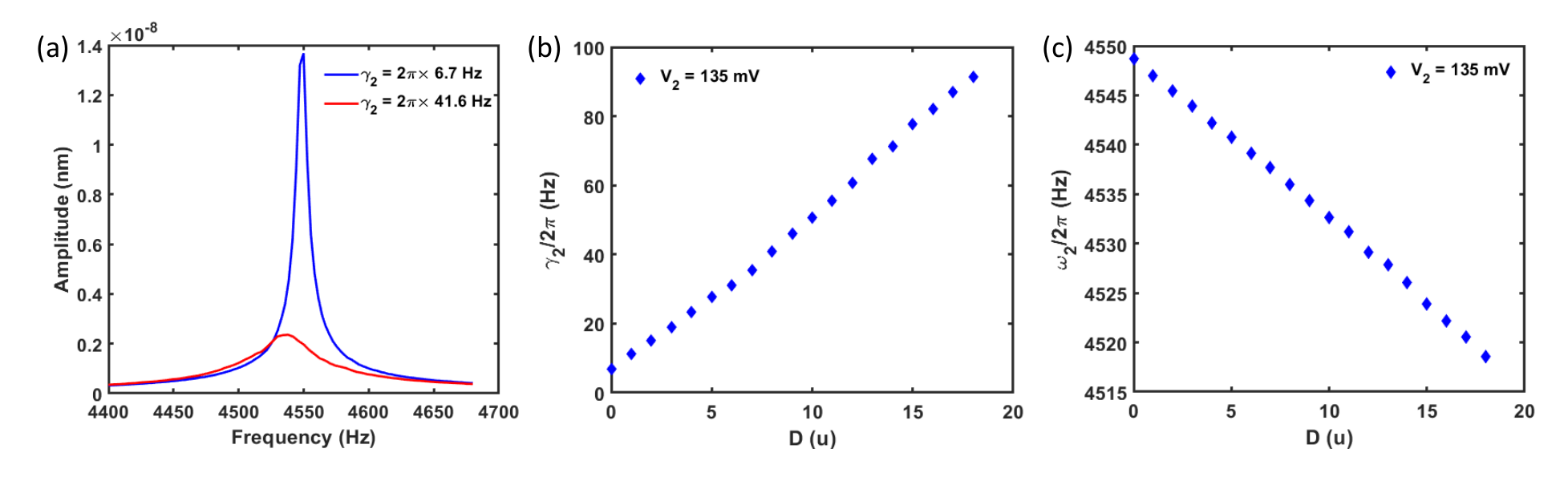}}
	\caption{Damping rate calibrated from the frequency response. (a). An example of the frequency response of cantilever 2 for two PID cases that $D = 0$ and $D = -8u$, which correspond to $\gamma_2 = 2\pi\times 6.7$ Hz and $\gamma_2 = 2\pi\times 41.6$ Hz. (b). The calibrated damping rate $\gamma_2$ is shown as a function of the externally controlled derivative parameters D. (c). The calibrated resonant frequency is shown as a function of the controlled derivative parameters D.  }
	\label{Damp_Fitting}
\end{figure}

In the experiment, we monitor the oscillation of two cantilevers by the fiber interferometers at each side. A PID feedback control loop is applied to cantilever 2 through the piezo chips to control the damping rate of cantilever 2 as shown in Fig.\ref{Scheme_setup}. 
We use the sweeper function in Zurich Instruments MFLI device to calibrate the damping rate of two cantilevers. The frequency response of a cantilever is fitted with the Lorentzian function to extract the damping rate. An example of the frequency response recorded by the sweeper function is shown in Fig.\ref{Damp_Fitting}.(a) and this corresponds to two special case of $D = 0 $ and $D = -8u$, where D is the derivative parameters in the PID control. 
From the Lorentzian fitting, we can get that $D = 0$ indicates a natural damping rate of $\gamma_2 = 2\pi\times 6.7$ Hz and $D = -8u$ gives a damping rate of $\gamma_2 = 2\pi\times 41.6$ Hz.  The calibrated damping rate and resonant frequency at different derivative parameters $D$ is shown in Fig.\ref{Damp_Fitting}.(b) and (c). Experimentally, we can tune the damping rate of cantilever 2 up to $2\pi\times 91$ Hz. 

\end{document}